# Morphology of Ti on Monolayer Nanocrystalline Graphene and its Unexpectedly Low Hydrogen Adsorption

Yuya Murata[1], Stefano Veronesi[1], Dongmok Whang[2], and Stefan Heun[1,*]

[1]*NEST, Istituto Nanoscienze-CNR and Scuola Normale Superiore, Piazza San Silvestro 12, 56127 Pisa, Italy*

[2] *School of Advanced Material Science and Engineering, SKKU Advanced Institute of Nanotechnology, Sungkyunkwan University, Suwon 16419, Korea.*

Abstract

Hydrogen adsorption on graphene can be increased by functionalization with Ti. This requires dispersing Ti islands on graphene as small and dense as possible, in order to increase the number of hydrogen adsorption sites per Ti atom. In this report, we investigate the morphology of Ti on nanocrystalline graphene and its hydrogen adsorption by scanning tunneling microscopy and thermal desorption spectroscopy, and compare the results with equivalent measurements on single-crystalline graphene. Nanocrystalline graphene consists of extremely small crystal grains of < 5 nm size. Ti atoms preferentially adsorb at the grain boundaries of nanocrystalline graphene and form smaller and denser islands compared to single-crystalline graphene. Surprisingly, however, hydrogen adsorbs less to Ti on nanocrystalline graphene than to Ti on single-crystalline graphene. In particular, hydrogen hardly chemisorbs to 1 ML of Ti on nanocrystalline graphene. This may be attributed to strong bonds between Ti and defects located along the grain boundaries in nanocrystalline graphene. This mechanism might apply to other metals, as well, and therefore our results suggest that when functionalizing graphene by metal atoms for the purpose of hydrogen storage or other chemical reactions, it is important to consider not only the morphology of the resulting surface, but also the influence of graphene on the electronic states of the metal.

---

[*] E-mail address: stefan.heun@nano.cnr.it

# 1. Introduction

The interest in metal-hydrogen interactions started some time ago, mainly driven by technological applications in many strategic fields such as fission and fusion reactors, fuel cells, and hydrogen storage. Metal hydrides have been extensively investigated in the sixties and seventies, exploring pure metals and metal alloys.[1] In particular, transition metals are appealing due to their high hydrogen storage capability both as pure metals and as alloys. Hydrogen is a promising energy carrier, since its combustion product is only water, which has triggered the first automotive applications of solid-state hydrogen storage devices back in 1974 (Billings Energy Corporation and Daimler-Benz Co.). The experimental vehicle presented by Daimler-Benz was fueled by a 200 kg Titanium-iron hydrogen tank,[1] allowing a range of 130 km at a speed of 60 km/h. The main drawbacks of such metal hydride devices are the elevated mass of the tank necessary to achieve a reasonably performance,[2] and the high cost of the raw materials needed to produce the storage devices.

A promising perspective to overcome these issues is to move towards metal nanoparticles, which can increase the storage capability, save weight, and cut costs. Indeed, there is evidence of an increased hydrogen uptake in polycrystalline metals with vacancies, dislocations, and grain boundaries.[3,4,5] This approach can be fruitfully implemented with new 2D materials providing the physical support for the nanostructured transition metals.

In this context, graphene has recently attracted attention as a storage material for hydrogen owing to its high surface area-to-mass ratio.[6] Hydrogen adsorption on graphene can be increased by functionalization with Ti.[7,8,9] When Ti atoms are deposited on pristine epitaxial monolayer single-crystalline graphene at room temperature, they diffuse and form islands with an average diameter of 10 nm.[10] Ti atoms inside such 3D islands are not expected to contribute to hydrogen adsorption, but only those at the surface contribute. It is therefore required to disperse the Ti on graphene in islands as small and dense as possible, in order to increase the number of hydrogen adsorption sites per Ti atom. Our group has demonstrated that when defects are introduced in graphene by ion bombardment, the diffusion of Ti atoms on graphene is reduced, and the average diameter of the Ti islands decreases to 5 nm.[11]

In this report, we investigate the morphology of Ti on nanocrystalline graphene and its hydrogen adsorption, and compare the results with equivalent measurements on single-crystalline graphene. Nanocrystalline graphene consists of small crystal grains of less than 5 nm in diameter.[12] Nanocrystalline graphene is obtained by carefully controlling nucleation during chemical vapor deposition (CVD) of graphene on Ge(110) surfaces. The grain boundaries consist of defects, and the distribution of these defects is better controlled than those introduced by ion bombardment. The morphologies of Ti on single-crystalline graphene and on nanocrystalline graphene are compared by scanning tunneling microscopy (STM). The hydrogen adsorption on Ti on these samples is evaluated by thermal desorption spectroscopy (TDS).

## 2. Experimental Section

The single-crystalline graphene and the nanocrystalline graphene films were grown by CVD on Ge(110) substrates.[12,13] The H-terminated Ge(110) substrate was loaded into a CVD chamber, and the vacuum level was maintained below ~$10^{-6}$ torr. To eliminate other sources of contamination, an epitaxial Ge layer was deposited right before graphene growth. Single-crystal monolayer graphene was grown by flowing a 100:1 mixture of $H_2$ and $CH_4$ gases at 80 torr and 920 °C for 2 h. For growth of nanocrystalline graphene, a 10:1 mixture of $H_2$ and $CH_4$ gases (10 torr) was introduced into the chamber at 900 °C for 1 h. During graphene growth with high $H_2/CH_4$ ratio, reversible etching and regrowth enable epitaxial growth of low-defect single-crystalline graphene on Ge substrate, because graphene defects are preferentially removed during etching reaction. When the $H_2$ partial pressure is lowered, however, the etching of defect is significantly suppressed, and thus nanocrystalline graphene with a large number of structural defects is obtained.

The samples were then introduced into an ultra-high vacuum chamber with a base pressure of $1\times10^{-10}$ mbar. The vacuum chamber is equipped with an electron beam evaporator for Ti (SPECS), a deuterium doser, a sample heating/cooling stage, a residual gas analyzer (Stanford Research Systems), an STM (RHK technology), and a low energy electron diffraction (LEED) system (OCI vacuum microengineering). All experiments were performed *in situ*, i.e., Ti evaporation, hydrogen exposure, STM, LEED, and TDS were performed in the same vacuum system, without exposing samples to air. The samples were degassed at 700 K for 14 hours in the vacuum chamber. Ti was deposited on the samples at room temperature. The deposition rate was 0.023 ML/s (1 ML = $1.32\times10^{15}$ atoms/cm$^2$). The samples were characterized by STM and LEED before and after Ti deposition.

After exposure of the samples to molecular deuterium, TDS was performed. Deuterium (mass 4) was used instead of hydrogen (mass 2) for a better signal-to-noise ratio in TDS. We refer to *hydrogen* and not *deuterium* throughout the paper, because both are chemically identical, except for a small shift in desorption temperature due to the isotope effect.[14] The samples were exposed to deuterium at 100 K for 5 min with a partial pressure of $3.5 \times 10^{-8}$ mbar. In TDS, while the samples were heated from 100 K to 700 K at a constant rate of about 6 K/s, the partial pressure of mass 4 was measured by the residual gas analyzer placed in front of the sample.

## 3. Results and Discussion

Figure 1 shows STM and LEED data obtained on single-crystalline graphene. Some wrinkles in graphene with a height of about 0.7 nm, which is twice the spacing between graphene layers, are indicated in Fig. 1a by arrows. It has been reported that graphene on Ge(110) has a compressive strain.[15] The wrinkles may be formed during cooling from the growth temperature, to relax the strain due to the difference of the thermal expansion coefficients between graphene and Ge. The high-resolution image of Fig. 1b shows the honeycomb lattice of graphene together with stripes with a periodicity of 0.57 nm, which corresponds to the Ge lattice constant along Ge(110)-[110].[16,17] The orientation relationship obtained from the STM image is graphene armchair // Ge(110)-[110], consistent with previous reports.[13,18] The LEED pattern in Fig. 1c shows spots corresponding to graphene and Ge. The orientation relationship between graphene and Ge obtained from the LEED data is the same as from the STM. Figure 1d shows an STM image from single-crystalline graphene after deposition of 0.55 ML Ti. It shows islands with a diameter of 5-10 nm. The islands cover 20-30% of the graphene surface. Their height corresponds to 4-9 atomic layers of Ti. The density of the islands is 0.4-0.5 per 100 $nm^2$. We found that some Ti islands nucleated on graphene defects, while others nucleated on graphene without a defect. A similar morphology of Ti islands has been observed on epitaxial monolayer single-crystalline graphene on SiC.[10]

Figure 2 shows the STM and LEED data from nanocrystalline graphene. Figure 2a shows that the nanocrystalline graphene consists of grains with a size of 3-5 nm. The higher-resolution image of Fig. 2c shows periodic structures in the grains. In the further magnified image in Fig. 2e, a periodic structure corresponding to a graphene $\sqrt{3} \times \sqrt{3}$ superstructure is observed in the grain. This $\sqrt{3} \times \sqrt{3}$ periodicity is characteristic of interference due to electrons scattered at the armchair edges of graphene.[19] This electron scattering at grain boundaries is consistent with the report that nanocrystalline graphene shows a high resistance in transport measurements.[12] The atomic structure near the grain boundary was not clearly resolved in the STM images in Figs. 2a and 2c. However, the line profile in Fig. 2b indicates a corrugation with amplitude of about 1 nm, and that the grain boundaries are lower than the center of the grains. It has been theoretically shown that the edge of graphene chemically bonds to the Ge substrate during the growth of graphene on Ge(110) terminated with hydrogen.[17] On the other hand, the interaction between the center of the graphene grains and the Ge substrate is weaker than at the edge. Strain in nanocrystalline graphene may be relaxed by the corrugation, instead of a wrinkling seen in the case of single-crystalline graphene. Note that the defect density in nanocrystalline graphene is much larger than the defect density in single-crystalline graphene.

In order to indicate the orientation of the grains more clearly, a derivative image of Fig. 2c was taken, which is shown in Fig. 2d. The green bars indicate the armchair direction for each grain. A strong variation in the orientation of the grains is evident. Figure 2f shows a histogram of the angular armchair orientation distribution of 86 grains. The angle was measured relative to the Ge(110)-[110] direction. Although there are grains observed in all directions, there is a weak preference. The full width at half maximum around the preferred angle is 12.5°. Consistent with this, the LEED pattern in Fig. 2g shows a streak along the azimuthal direction originating from the graphene lattice. The corresponding line profile of the streak along the azimuthal direction is shown in Fig. 2i, taken along the blue line in Fig. 2g. The full width at half maximum of the streak is 16.8°. The angle with the maximum intensity corresponds to the Ge(110)-[110] direction, as shown in another LEED pattern taken at different electron energy (Fig. 2h). This orientation relationship between graphene and Ge is the same as for single-crystalline graphene. Thus, this orientation seems to be the most stable energetically.

Next, nanocrystalline graphene after deposition of 0.55 ML Ti was observed by STM. Figure 3 shows the results. With a tunneling current of $I_t$ = 0.03 nA and sample bias voltage $V_b$ in the range from 0.1 V to 1.6 V, the observation was unstable. With $V_b$ = 4-5 V, the observation was stable, and Ti islands could be observed. Finally, with $V_b$ between 2 V and 3 V, after a few unstable initial scans, a surface morphology similar to that of clean nanocrystalline graphene appeared. This suggests that scans with $V_b$ = 2-3 V remove Ti. Figure 3a shows an STM image of nanocrystalline graphene after deposition of 0.55 ML Ti, obtained with $V_b$ = 4 V. Before this observation, the area on the right side of the green line had been already scanned a few times with $V_b$ = 2 V. The morphology of the right area is relatively smooth and similar to that before Ti deposition. On the other hand, on the left side, there is a dense population of small islands. After this measurement, the whole area of Fig. 3a was scanned a few times with $V_b$ = 2 V, then again observed with $V_b$ = 4 V. The result is shown in Fig. 3b. The right side of the image did not change, but the islands on the left side disappeared, and consequently the whole area now seems uniform. We have verified that the perimeter of areas scanned with $V_b$ = 2 V did not show any changes, which indicates that Ti has moved to the STM tip instead of moving laterally on the surface. The removal of metal atoms from graphene by an STM tip has been reported by other groups before.[20,21] These groups have reported that the efficiency of the removal of metal atom is higher when the distance between the STM tip and the sample is shorter. For a given tunneling current, this distance generally becomes shorter with smaller $V_b$. This is consistent with our observations. We add that also Ti on single-crystalline graphene was removed by the STM tip when working with a bias of $V_b$ = 1-3 V, but on the other hand, the observation was relatively stable with $V_b$ = 0.1-0.5 V, as witnessed by the image shown in Fig. 1d, and this is in contrast with the case of nanocrystalline graphene. We will later discuss the different interactions between Ti and the two types of graphene samples.

STM images on nanocrystalline graphene after Ti deposition reflect both the morphology of the nanocrystalline graphene and the Ti islands. This allows subtracting the image after Ti removal from the image before Ti removal, in order to obtain more precise information about the morphology and distribution of the Ti islands. Figure 3c shows the subtraction image of Fig. 3b from Fig. 3a. The right side is flat because there the Ti islands had already been removed in Fig. 3a, so the scans between Figs. 3a and 3b with $V_b$ = 2 V did not cause much change. The left side, however, shows the morphology of only the Ti islands. It shows islands with a size of 3-5 nm. The surface is covered to 73% by Ti. The height of the islands is 0.28 nm, which corresponds to 1.2 atomic layers of Ti. The density of the islands is 7-8 per 100 $nm^2$. Furthermore, from a comparison between Figs. 3b and 3c, we determine that the Ti prefers to adsorb on the grain boundaries of the nanocrystalline graphene and not on the grains. For this purpose, we first take a histogram of Fig. 3b, which reflects the height distribution of the graphene surface. Next, we compare this to the height distribution of that part of the surface which was covered by Ti. For this purpose, we calculate a threshold image of Fig. 3c in which we set areas with Ti to 1 and areas without Ti to 0. The result is shown in Fig. 3d. Multiplying this image with the data of Fig. 3b gives an image in which all parts of the surface which were *not* covered by Ti have been removed by setting them to zero. The resulting height distribution histogram is shown in Fig. 3e, green curve. Following the same procedure, we also calculate the height distribution of the part of the surface which was *not* covered by Ti, shown in Fig. 3e, red curve (by using the inverse of Fig. 3d). Clearly, the green histogram is centered at lower height values as compared to the red histogram. This indicates that the Ti grows preferentially in the lower regions of the nanocrystalline graphene surface, and these are, as we have already shown, the positions of the grain boundaries. As a crosscheck, we have added the green and red histograms, and, as expected, their sum equals the histogram of the whole image of Fig. 3b. After Ti deposition, the nanocrystalline graphene sample was annealed at 800 K for 14 hours. The morphology of the sample did not change significantly by the annealing.

Figure 4 shows the surface coverage and the surface area of Ti islands for 0.55, 2, and 3 ML. Both the surface coverage and the surface area of Ti islands on nanocrystalline graphene are larger than those on single-crystalline graphene. The difference is larger when the deposited Ti amount is smaller. We conclude that smaller and denser Ti islands and higher surface coverage are obtained on nanocrystalline graphene as compared to single-crystalline graphene. On nanocrystalline graphene, Ti preferentially adsorbs on grain boundaries. Smaller Ti islands are formed because diffusion of Ti atoms is suppressed due to strong bonds between Ti and defects at grain boundaries in nanocrystalline graphene.

The line profiles in Figs. 4c and 4d show that the morphology of nanocrystalline graphene after Ti deposition of 0.55 and 2 ML, respectively, is similar to that before Ti deposition (cf. Fig. 2b), indicating layer-by-layer growth of the Ti. This was also shown by comparison of STM images before and after removal of Ti in Fig. 3. Therefore, after deposition of 2 ML of Ti, the top layer, i.e., the second layer of Ti atoms, is not in contact with graphene. On the other hand, Ti of 0.55 and 2 ML deposited on single-crystalline graphene forms 3D islands, as seen in Figs. 4e and 4f, respectively. Thus, with 2 ML deposited, most of the Ti surface atoms are not in contact with graphene for both types of graphene.

Next, we evaluated the hydrogen adsorption on these samples by TDS. Figure 5 shows the TDS results on the single-crystalline graphene and the nanocrystalline graphene. Hydrogen desorption was not seen from the samples without Ti (black curves). Desorption spectra from samples with Ti showed peaks at around 200 K and 480 K. The corresponding desorption energies were estimated from the desorption temperatures and are shown in table 1. For this purpose, we assumed first-order desorption and a typical attempt frequency of $10^{13}$ Hz.[10] The error bars correspond to the full width at half maximum of the desorption peaks. Desorption energies around 0.5 and 1.3 eV, estimated from the desorption peaks at 200 K and 480 K, respectively, are close to those obtained for Ti on graphene on SiC.[22] Therefore, the desorption peaks at 200 K and 480 K can be attributed to hydrogen desorption related to physisorption and chemisorption, respectively. Here, physisorption refers to adsorption of hydrogen molecules, whereas chemisorption to adsorption of hydrogen atoms.

In order to estimate the amount of adsorbed hydrogen, the partial pressure of hydrogen in the TDS spectra was integrated under each desorption peak, and is plotted as a function of Ti amount in Fig. 5c. The peaks in the TDS spectra were fitted based on the model reported previously,[23] and then they were integrated with respect to time. Figure 5c shows that the total amount of adsorbed hydrogen on nanocrystalline graphene is smaller than that on single-crystalline graphene. Particularly, with 1 ML of Ti, a clear peak related to chemisorption was observed on single-crystalline graphene, but it was negligibly small on nanocrystalline graphene. Theoretical calculation has reported that Ti bonding to defects in graphene does not chemisorb hydrogen.[22] This was explained by a strong bond between the Ti and a defect in graphene mediated by charge transfer. This and the TDS results suggest that most of the 1 ML Ti adsorbs near defects of nanocrystalline graphene, which inhibits chemisorption of hydrogen due to charge transfer. We caution the reader that we have no direct spectroscopic signature of this charge transfer. We performed scanning tunneling spectroscopy on Ti islands deposited on the two different graphene samples, to measure their electronic states. However, we could not resolve differences, probably due to an insufficient energy resolution of the measurements at room temperature. In another theoretical report, it has been shown that at room temperature Ti forms bonds to di-interstitial defects in curved graphene, but not on flat graphene without defects.[24] This is consistent with the differences in chemisorption between single-crystalline graphene and nanocrystalline graphene in the TDS data. Furthermore, the Ti-C bond on nanocrystalline graphene may explain the instability in STM at low $V_b$. TiC nanocrystals have a band gap of 0.8 eV.[25] If Ti forms bonds to defects in nanocrystalline graphene and a band gap is induced, a stable tunneling contact cannot be established for $V_b$ within the band gap, and thus STM imaging becomes unstable.

On the other hand, 1 ML of Ti on nanocrystalline graphene shows a higher physisorption peak than single-crystalline graphene. Physisorption does not involve a charge transfer between hydrogen and Ti.[22] Therefore, defects in nanocrystalline graphene may have less influence on physisorption than on chemisorption. As shown in Fig. 4b, Ti on nanocrystalline graphene has a larger surface area than on single-crystalline graphene, which leads to higher amount of physisorbed molecules. Increasing the Ti amount from 1 ML to 2 ML, a peak corresponding to chemisorption appeared in the TDS data on nanocrystalline graphene. This indicates that the second layer of Ti is less influenced by defects in nanocrystalline graphene, and so it can chemisorb hydrogen. However, chemisorption on single-crystalline graphene also increased with Ti amount, and it is still larger than that on nanocrystalline graphene. This suggests that the influence of defect on the second layer of Ti decreases but still exists.

## 4. Conclusions

We investigated the morphologies of Ti on single-crystalline graphene and nanocrystalline graphene by STM, and their hydrogen adsorption properties by TDS. STM revealed that smaller and denser Ti islands, and higher coverage of them, are formed on nanocrystalline graphene as compared to single-crystalline graphene. On nanocrystalline graphene, Ti preferentially adsorbs on grain boundaries. This suggests that strong bonds between Ti and the defects in nanocrystalline graphene suppress diffusion of Ti and lead to the formation of smaller islands. However, the TDS data showed that the amount of hydrogen adsorption is not simply proportional to the surface area of Ti. The total amount of adsorbed hydrogen on nanocrystalline graphene is smaller than that on single-crystalline graphene. In particular, hydrogen hardly chemisorbs to 1 ML of Ti on nanocrystalline graphene. This may be attributed to strong bonds between Ti and defects in nanocrystalline graphene. These results suggest that when functionalizing graphene by metal atoms for the purpose of hydrogenation or any other chemical reaction, it is essential to consider not only the morphology of the resulting surface, but also the influence of graphene on the electronic states of the metal. On the other hand, it might be possible to exploit the modification of the electronic states of deposited islands towards, for example, applications in optics.

**Acknowledgements**

We would like to thank the organizers of the Korea-Italy Bilateral Symposium on "Beyond Graphene", Hanyang University, Seoul, Korea, 27 May 2016. Financial support from CNR in the framework of the agreement on scientific collaboration between CNR and NRF (Korea) is acknowledged. We gratefully acknowledge financial support by EU-H2020 Graphene-Core2 (agreement No. 725219). DW acknowledges support from the National Research Foundation of Korea (NRF) grant funded by the Korea government (MSIP) (NRF-2017R1A2B2010663). We further acknowledge funding from the EC through the project PHOSFUN Phosphorene functionalization: a new platform for advanced multifunctional materials (ERC ADVANCED GRANT No. 670173). SH thanks Scuola Normale Superiore for support, project SNS16_B_HEUN—004155, and the Italian Ministry of Foreign Affairs, Direzione Generale per la Promozione del Sistema Paese (agreement on scientific collaboration between Italy and Poland).

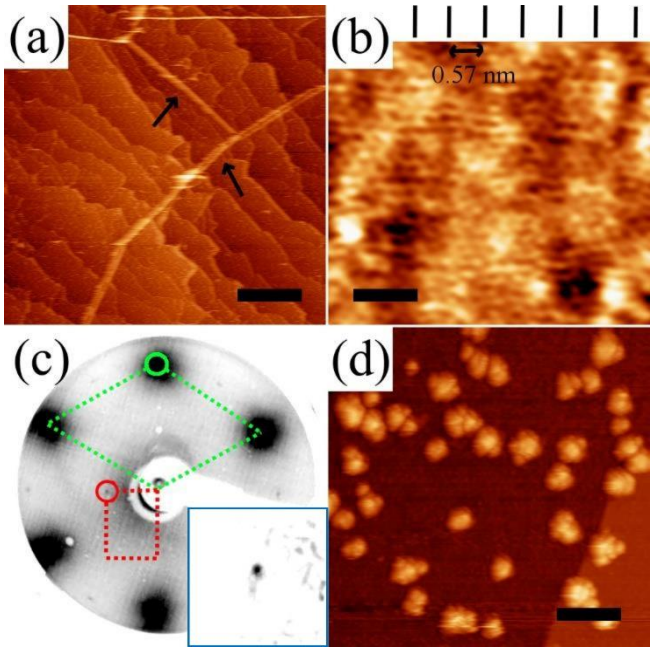

Fig. 1 (a) STM image of single-crystalline graphene. Sample bias voltage $V_b$ = -0.1 V, tunneling current $I_t$ = 0.035 nA, scan size = 1000 nm × 1000 nm. Wrinkles of graphene are indicated by arrows. Scale bar 200 nm. (b) STM image of single-crystalline graphene. $V_b$ = 0.2 V, $I_t$ = 0.1 nA, scan size = 5 nm × 5 nm. Vertical lines indicate the periodicity of the Ge(110)-[110] lattice (0.57 nm). Scale bar 1 nm. (c) LEED of single-crystalline graphene. Electron energy = 75 eV. The red and green circles indicate Ge(10) and graphene(11) spots, respectively. The dashed lines indicate the corresponding reciprocal unit cells. The inset shows a magnified view of the Ge(10) spot, background-subtracted and with maximized contrast. (d) STM image of single-crystalline graphene after Ti deposition of 0.55 ML. $V_b$ = -0.5 V, $I_t$ = 0.1 nA, scan size = 100 nm × 100 nm. Scale bar 20 nm.

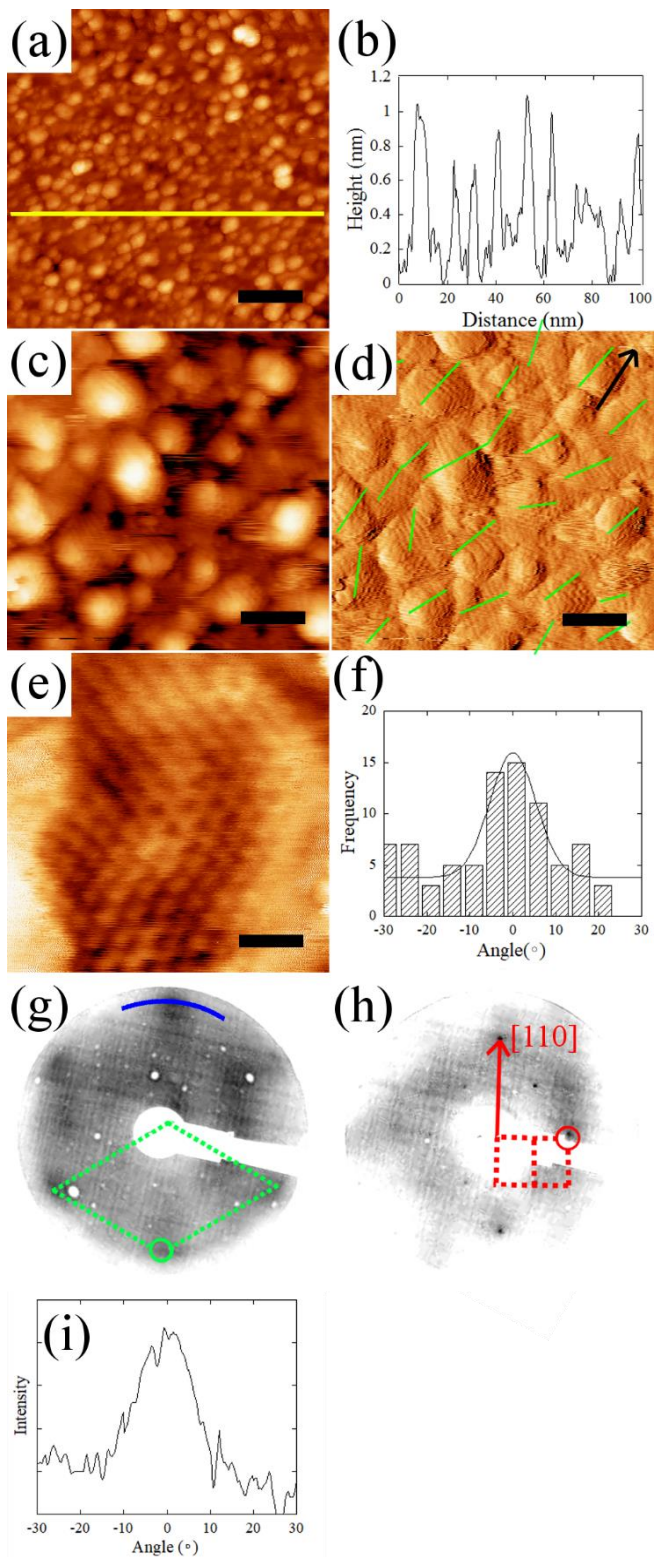

Fig. 2 (a) STM image of nanocrystalline graphene. $V_b$ = 0.2 V, $I_t$ = 0.2 nA, scan size = 100 nm × 100 nm. Scale bar 20 nm. (b) Line profile along the yellow line in (a). (c) STM image of nanocrystalline graphene. $V_b$ = -0.1 V, $I_t$ = 0.01 nA, scan size = 20 nm × 20 nm. Scale bar 4 nm. (d) same image as (c), but differentiated along the *x*-axis. The green bars indicate the armchair directions of the individual graphene domains. The black arrow indicates the Ge[110] direction. Scale bar 4 nm. (e) STM image of nanocrystalline graphene. $V_b$ = 0.1 V, $I_t$ = 0.2 nA, scan size = 5 nm × 5 nm. Scale bar 1 nm. (f) Histogram of the angular distribution of the armchair directions of 86 graphene domains. The angle is measured relative to the Ge[110] direction. (g) LEED of nanocrystalline graphene. Electron energy = 74 eV. The green circle indicates a graphene(11) spot, while the dashed green line indicates the corresponding reciprocal unit cell. (h) LEED of nanocrystalline graphene. Electron energy = 150 eV. The red circle indicates a Ge(20) spot. (i) Line profile of the graphene streak indicated by the blue arc in (g).

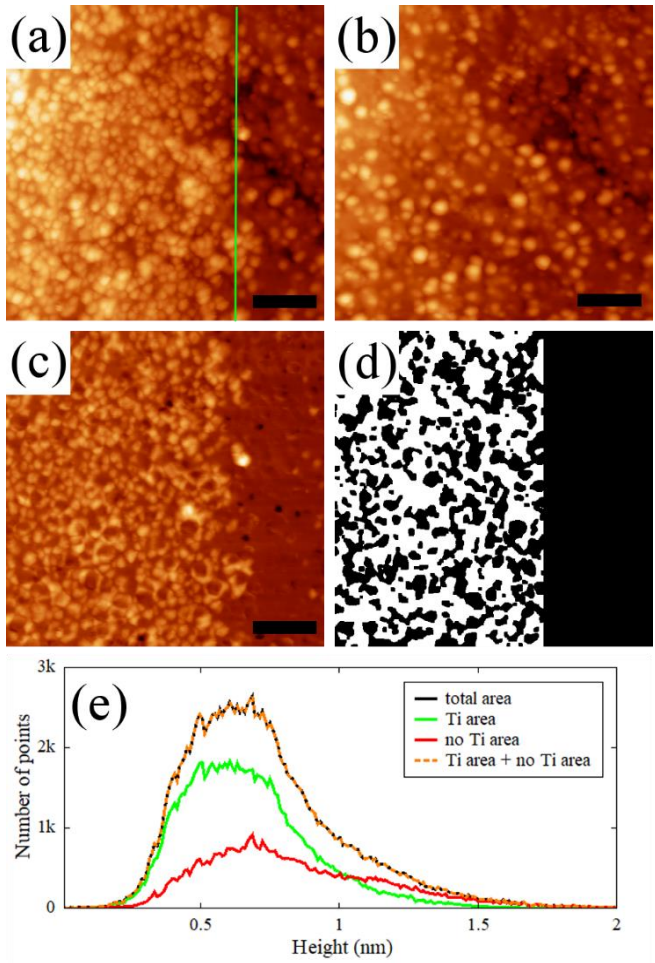

Fig. 3 (a) STM image of nanocrystalline graphene after 0.55 ML Ti deposition. $V_b$ = 4 V, $I_t$ = 0.03 nA, scan size = 100 nm × 100 nm. Scale bar 20 nm. (b) STM image taken from the same area as (a), after scanning several times with $V_b$ = 2 V and $I_t$ = 0.03 nA. Image parameters: $V_b$ = 4 V, $I_t$ = 0.03 nA, scan size = 100 nm × 100 nm. Scale bar 20 nm. (c) Difference image obtained by subtraction of (b) from (a). Scale bar 20 nm. (d) Threshold image of (c) in which areas with Ti are set to 1 and areas without Ti to 0. (e) Height histogram of (b). Black: from the total area, green: counting only pixels where Ti adsorbed, red: counting only pixels where Ti did *not* adsorb, orange dash: sum of the green and red curves.

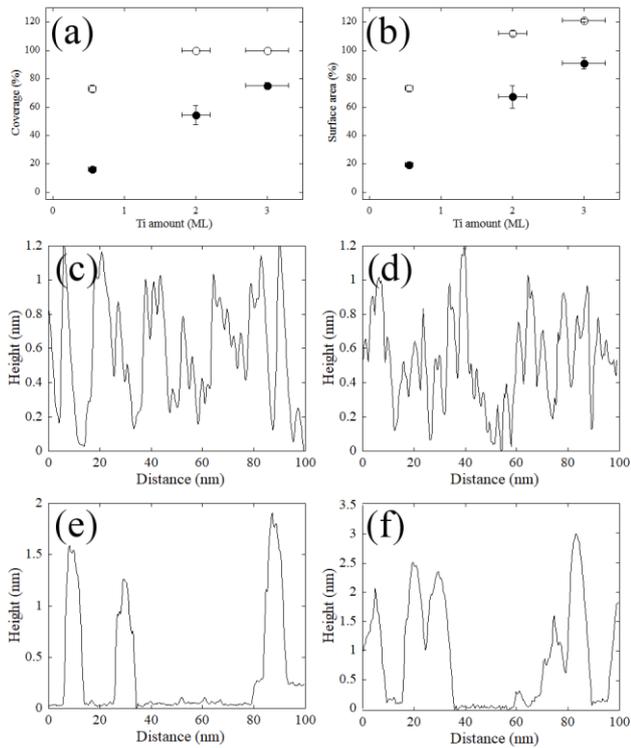

Fig. 4 (a) Surface coverage and (b) surface area of Ti islands per substrate area as a function of Ti amount on (hollow circle) nanocrystalline graphene and (filled circle) single-crystalline graphene. (c) and (d) Line profiles of nanocrystalline graphene after Ti deposition of 0.55 ML and 2 ML, respectively. (e) and (f) Line profiles of single-crystalline graphene after Ti deposition of 0.55 ML and 2 ML, respectively.

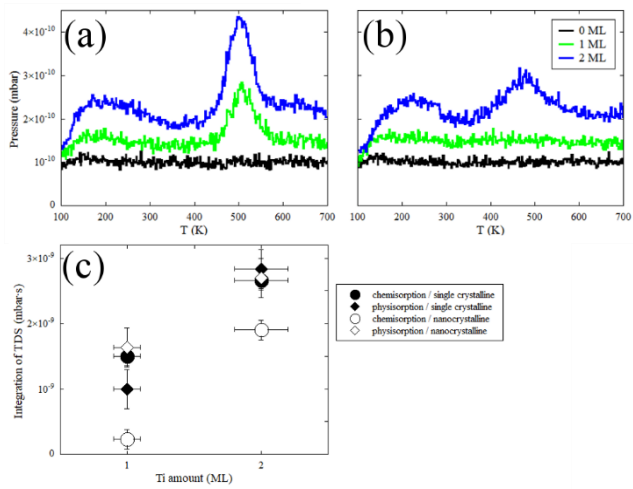

Fig. 5 TDS on (a) single-crystalline graphene and (b) on nanocrystalline graphene, with Ti of (black) 0 ML, (green) 1 ML, and (blue) 2 ML. (c) Integration of partial pressure of hydrogen in TDS data in the desorption peaks related to chemisorption and physisorption, plotted as a function of Ti amount.

Table 1 Desorption energies of hydrogen estimated from the peaks at low (200 K) and high temperature (480 K) in the TDS spectra taken on single-crystalline graphene and on nanocrystalline graphene, for 1 ML and 2 ML of Ti (in eV).

|      | single-crystal graphene | | nanocrystalline graphene | |
| --- | --- | --- | --- | --- |
|      | 200 K | 480 K | 200 K | 480 K |
| 1 ML | 0.44±0.24 | 1.35±0.1 | 0.49±0.26 | peak too small |
| 2 ML | 0.50±0.22 | 1.32±0.09 | 0.65±0.24 | 1.25±0.18 |

**References**

[1] *Hydrogen in Metals I and II*; Alefeld, G., Völkl, J., Eds.; Springer: Berlin, 1978.

[2] Tozzini, V.; Pellegrini, V. Prospects for Hydrogen Storage in Graphene. *Phys. Chem. Chem. Phys.* **2013**, *15*, 80 – 89.

[3] Clarebrough, L. M.; Humble, P.; Loretto, M. H. Voids in Quenched Copper, Silver and Gold. *Acta Met.* **1967**, *15*, 1007 – 1023.

[4] Wriedt, H. A.; Oriani, R. A. Effect of Tensile and Compressive Elastic Stress on Equilibrium Hydrogen Solubility in a Solid. *Acta Met.* **1970**, *18*, 753 – 760.

[5] da Silva, J. R. G.; Stafford, S. W.; McLellan, R. B. The Thermodynamics of the Hydrogen-Iron System. *J. Less-Common Metals* **1976**, *49*, 407 – 420.

[6] Sofo, J. O.; Chaudhari, A. S.; Barber, G. D. Graphane: A Two-Dimensional Hydrocarbon. *Phys. Rev. B* **2007**, *75*, 153401.

[7] Durgun, E.; Ciraci, S.; Yildirim, T. Functionalization of Carbon-Based Nanostructures with Light Transition-Metal Atoms for Hydrogen Storage. *Phys. Rev. B* **2008**, *77*, 085405.

[8] Liu, Y.; Ren, L.; He, Y.; Cheng, H.-P. Titanium-Decorated Graphene for High-Capacity Hydrogen Storage Studied by Density Functional Simulations. *J. Phys.: Condens. Matter* **2010**, *22*, 445301.

[9] Kim, G.; Jhi, S.-H.; Park, N.; Louie, S. G.; Cohen, M. L. Optimization of Metal Dispersion in Doped Graphitic Materials for Hydrogen Storage. *Phys. Rev. B* **2008**, *78*, 085408.

[10] Mashoff, T.; Takamura, M.; Tanabe, S.; Hibino, H.; Beltram, F.; Heun, S. Hydrogen Storage with Titanium-Functionalized Graphene. *Appl. Phys. Lett.* **2013**, *103*, 013903.

[11] Mashoff, T.; Convertino, D.; Miseikis, V.; Coletti, C.; Piazza, V.; Tozzini, V.; Beltram, F.; Heun, S. Increasing the Active Surface of Titanium Islands on Graphene by Nitrogen Sputtering. *Appl. Phys. Lett.* **2015**, *106*, 083901.

[12] Joo, W.-J.; Lee, J.-H.; Jang, Y.; Kang, S.-G.; Kwon, Y.-N.; Chung, J.; Lee, S.; Kim, C.; Kim, T.-H.; Yang, C.-W. et al. Realization of Continuous Zachariasen Carbon Monolayer. *Sci. Adv.* **2017**, *3*, e1601821.

[13] Lee, J.-H.; Lee, E. K.; Joo, W.-J.; Jang, Y.; Kim, B.-S.; Lim, J. Y.; Choi, S.-H.; Ahn, S. J.; Ahn, J. R.; Park, M.-H. et al. Wafer-Scale Growth of Single-Crystal Monolayer Graphene on Reusable Hydrogen-Terminated Germanium. *Science* **2014**, *344*, 286 – 289.

[14] Zecho, T.; Güttler, A.; Sha, X.; Jackson, B.; Küppers, J. Adsorption of Hydrogen and Deuterium Atoms on the (0001) Graphite Surface. *J. Chem. Phys.* **2002**, *117*, 8486 – 8492.

[15] Kiraly, B.; Jacobberger, R. M.; Mannix, A. J.; Campbell, G. P.; Bedzyk, M. J.; Arnold, M. S.; Hersam, M. C.; Guisinger, N. P. Electronic and Mechanical Properties of Graphene−Germanium Interfaces Grown by Chemical Vapor Deposition. *Nano Lett.* **2015**, *15*, 7414 – 7420.

[16] Cooper, A. S. Precise Lattice Constants of Germanium, Aluminum, Gallium Arsenide, Uranium, Sulphur, Quartz and Sapphire. *Acta. Cryst.* **1962**, *15*, 578 – 582.

[17] Dai, J.; Wang, D.; Zhang, M.; Niu, T.; Li, A.; Ye, M.; Qiao, S.; Ding, G.; Xie, X.; Wang, Y. et al. How Graphene Islands Are Unidirectionally Aligned on the Ge(110) Surface. *Nano Lett.* **2016**, *16*, 3160 – 3165.

[18] Rogge, P. C.; Foster, M. E.; Wofford, J. M.; McCarty, K. F.; Bartelt, N. C.; Dubon, O. D. On the Rotational Alignment of Graphene Domains Grown on Ge(110) and Ge(111). *MRS Communictions* **2015**, *5*, 539 – 546.

[19] Sakai, K. I.; Takai, K.; Fukui, K. I.; Nakanishi, T.; Enoki, T. Honeycomb Superperiodic Pattern and its Fine Structure near the Armchair Edge of Graphene Observed by Low-Temperature Scanning Tunneling Microscopy. *Phys. Rev. B* **2010**, *81*, 235417.

[20] Donner, K.; Jakob, P. Structural Properties and Site Specific Interactions of Pt with the Graphene/Ru(0001) Moiré Overlayer. *J. Chem. Phys.* **2009**, *131*, 164701.

[21] Martínez-Galera, A. J.; Brihuega, I.; Gutiérrez-Rubio, A.; Stauber, T.; Gómez-Rodríguez, J. M. Towards Scalable Nano-Engineering of Graphene, *Sci. Rep.* **2014**, *4*, 7314.

[22] Takahashi, K.; Isobe, S.; Omori, K.; Mashoff, T.; Convertino, D.; Miseikis, V.; Coletti, C.; Tozzini, V.; Heun, S. Revealing the Multibonding State between Hydrogen and Graphene-Supported Ti Clusters. *J. Phys. Chem. C* **2016**, *120*, 12974 – 12797.

[23] Chen, R. Evaluation of Parameters from Thermal Desorption Spectra – Methods Borrowed from the Analysis of Thermoluminescence. *Surf. Sci.* **1998**, *400*, 258 – 265.

[24] Fonseca, A. F.; Liang, T.; Zhang, D.; Choudhary, K.; Phillpot, S. R.; Sinnott, S. B. Titanium-Carbide Formation at Defective Curved Graphene-Titanium Interfaces. *MRS Advances* **2018**, *3*, 457 – 462.

[25] Enyashin, A. N.; Ivanovskii, A. L. Electronic Structure of Extended Titanium Carbide Nanocrystallites. *J. Structural Chem.* **2006**, *47*, 549 – 552.